\documentclass{iopart}

\usepackage{amsmath} 
\usepackage{amssymb} 
\usepackage{graphicx}
\usepackage{xspace}
\newcommand{\p}{\partial} 
\newcommand{\pdt}{\partial_t} 
\newcommand{\tp}{\tilde{\psi}}
\newcommand{\hphi}{\hat{\phi}}
\newcommand{\tphi}{\tilde{\phi}}
\newcommand{\hp}{\hat{\psi}}

\newcommand{\tj}{\hat{j}}
\newcommand{\gk}{\Gamma_k}
\newcommand{\bq}{{q}}
\newcommand{\bx}{{x}}

\newcommand{\anz}{{\em ansatz}\xspace}
\newcommand{\nequ}{non-equilibrium\xspace}
\newcommand{\npt}{non-perturbative\xspace}

\begin{document}

\title{Reaction-diffusion processes and \npt renormalisation group}
\author{L\'eonie Canet}
\address{Theory Group, School of Physics and Astronomy, University of Manchester, Manchester M13 9PL, United Kingdom.}





\begin{abstract}
This paper is devoted to investigating \nequ phase transitions to an absorbing
 state, which are  generically  encountered in reaction-diffusion processes.
 It is a review, based on \cite{canet04a,canet04b,canet05}, of recent progress in this field
  that has been allowed by  a \npt renormalisation group approach. 
 We mainly focus on branching and annihilating
 random walks and show that  their critical properties  strongly rely on 
  \npt features and that hence the use of a \npt method 
  turns out to be crucial to get a correct picture of the physics of these models.

\end{abstract}

\date{\today}

\section{Introduction}
Our understanding of equilibrium critical phenomena has largely benefited from the success of perturbative 
renormalisation group (RG) techniques and from conformal symmetry properties in two dimensions.
  Unfortunately, this success story has not spread out to 
 \nequ systems, though the latter are far more common in nature.
  Indeed,  the study of phase transitions between \nequ steady states has taught us that critical 
  phenomena or generic scale invariance turn out to be much richer far from thermal equilibrium ---
  where detailed balance relations
  are violated --- than in equilibrium statics or even near-equilibrium dynamics. 
 
 However, despite considerable efforts, the very ingredients settling 
 the universality classes out of equilibrium remain on a fragile footing.  
 The problem originates mostly in the absence 
  of an effective free energy functional that would allow to straightforwardly
  classify the universal behaviours  in terms of
  symmetries and interactions. And even when such a functional exists --- which occurs 
   for reaction-diffusion processes or for  Langevin type dynamics ---  analytical
   progress turns out to be difficult.
  On the one hand, models with Langevin dynamics cannot be conformal invariant. 
   On the other hand, the efficiency of standard
  RG approaches is hindered by the fact that first, critical dimensions,
   when they are identified, happen
 to lie far from the physically interesting ones and furthermore, no calculations are available above two-loop
   order, which prevent from resorting to powerful re-summation techniques.
  In this context,
  any theoretical tool that can overcome the previous pitfalls
  is valuable and the \npt renormalisation group (NPRG) stands as 
 a promising candidate \cite{bagnuls01}. Indeed, this method has lead to great successes for systems 
 at equilibrium during the last decade, appearing as a well-adapted tool to tackle strong coupling
  problems \cite{bagnuls01,berges02}. Moreover,
  the  NPRG formalism has been recently extended to \nequ systems \cite{canet04a} and 
 has given rise to important progress in the field of reaction-diffusion processes,
 unveiling genuinely \npt effects \cite{canet04a,canet04b,canet05}, which we review in this paper.
 
 This paper is organised as follows. In section \ref{secII}, we give an overview 
 of reaction-diffusion processes and recall their  field theoretical formulation. 
 In section \ref{secIII}, we  briefly introduce the NPRG formalism generalised to \nequ systems and
  we  derive the flow equations related to reaction-diffusion processes, before focusing on specific
 models. Some results on the universal
  properties of directed percolation (DP) are first reviewed  in section \ref{secIV} and
   the remainder of the
  paper (section \ref{secV})
 is devoted to the study of branching and annihilating random 
 walks (BARW), which split into two categories, namely the `odd' and `even' BARW.

\section{Reaction-diffusion processes}
\label{secII}

Reaction-diffusion processes constitute simple models that allow to gain some insights in \nequ critical
 phenomena \cite{hinrichsen00}. They are defined by a set of microscopic rules that govern
 the dynamics of a species of particles $A$. These particles 
   randomly diffuse at a rate $D$
 and undergo some reactions, typically birth $A\xrightarrow{\sigma_m} (m+1)A$ and death $kA\xrightarrow{\lambda_k} \varnothing$ processes. 
  From these competing interactions emerges at long time a stationary state which nature depends
 on the reaction rates $\lambda_k$ and $\sigma_m$. Either all particles eventually vanish, leaving 
 an empty state where all stochastic fluctuations cease and which is therefore called `absorbing'. 
 Or the density eventually saturates to 
 a finite average value yielding an `active' state where the dynamics constantly generates
     density fluctuations. 
  The active and  absorbing states are separated by a continuous phase transition \cite{hinrichsen00}.
 
 A large range of the absorbing transitions encountered in reaction-diffusion processes fall into the 
 DP universality class \cite{stauffer92}, which stands as the most prominent one. It has lead to a famous 
  conjecture in the early eighties by Janssen and Grassberger \cite{janssen81} stating that 
   a continuous transition to an absorbing state driven by a one-component order parameter 
 will generically fall into the DP universality class (provided there is no additional symmetry or
 quenched disorder).
 The DP model, which corresponds to the rules
\begin{equation}
 A \xrightarrow{\sigma} 2A, \hspace{1cm} 2A \xrightarrow{\lambda}\varnothing,
 \hspace{1cm} A \xrightarrow{\mu}\varnothing,
\label{DPrules}
\end{equation}
 therefore 
 plays a paradigmatic role, as the counterpart of the Ising model for equilibrium statistical physics.\\

\begin{figure}[ht]
\caption{Time evolution of particles on a one-dimensional lattice, starting from either 
a uniformly distributed initial configuration (top row) or a seed particle (bottom row), 
 for the DP model (endowed with the dynamical rules (\ref{DPrules})) from \cite{hinrichsen00}.
 In this model, the stationary states (respectively absorbing on the left-hand side 
 and active on the right-hand side) are reached exponentially fast in time, 
 and are separated by a critical state (middle column) where both time and space correlation lengths diverge,
  resulting in  an algebraic decay of the density.\\}
\begin{indented}
\item[]\includegraphics[width=75mm,height=55mm]{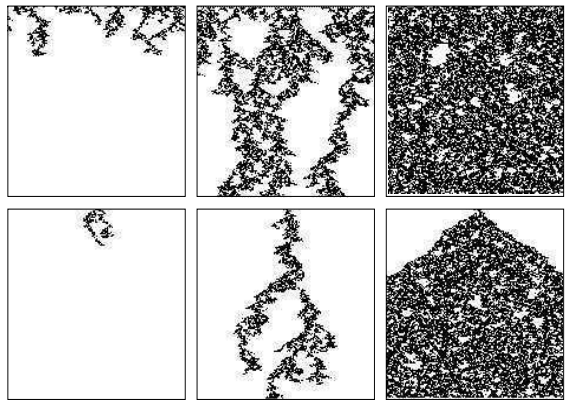}
\end{indented}
\label{percoh}
\end{figure}

 Reaction-diffusion processes  naturally lend themselves to Monte-Carlo simulations, 
 which have indeed largely contributed to
 our understanding of these processes (see \cite{hinrichsen00,odor04} for reviews). 
  On the other hand, the simplest analytical approach is to device a rate equation
 for the time-dependent  average 
 density $n(t)$, assuming  the various reactions to  
 occur proportionally to the concentration of reactants.
  For instance, for the DP processes (\ref{DPrules}), this yields:
\begin{equation}
\pdt \, n(t) = (\sigma - \mu) \,n(t) - 2\, \lambda \,n(t)^2. \label{law}
\end{equation}
 Eq. (\ref{law}) entails a mean-field type of approximation since  
 the density correlations are neglected
 ---  the joint probability of finding two particles at the
 same position has been factored. Eq. (\ref{law}) is readily solved. It
 exhibits two stationary solutions $n_{a}=0$ and $n_{s}= (\sigma - \mu)/(2\,\lambda)$, which stability
 depends on the sign of the `mass' $\Delta = (\sigma - \mu)$. If $\Delta < 0$, the density decreases
 to the absorbing solution $n_{a}$, whereas it saturates to 
 the active solution $n_{s}$ for $\Delta > 0$. 
 The explicit solution for the time-dependent density $n(t) = n_0 n_s/[n_0 + (n_s-n_0)\exp(-\Delta t)]$
  shows that both asymptotic states are reached exponentially fast in time.
  The relaxation time $\Delta^{-1}$  diverges when $\sigma=\mu$ causing an algebraic decay
 of the density 
 which corresponds to the critical state.
 The DP absorbing phase transition is illustrated in Fig. \ref{percoh} \cite{hinrichsen00}, which represents 
 the  time evolution of particles on a one-dimensional lattice for increasing values of $\Delta$ (see caption). 

 Absorbing phase transitions can be characterised by a set of critical exponents,
   typically:
\begin{eqnarray}
 n_s &\sim& (p-p_c)^{\beta} \label{defexpobeta}\\
 \xi_\perp &\sim& |p-p_c|^{ -\nu_\perp} \label{defexponuper}\\
 \xi_\| &\sim& |p-p_c|^{- \nu_\|}, \label{defexponupar}
\end{eqnarray}
where  $z=\nu_\|/\nu_\perp$ embodies the dynamical exponent, which represents the anomalous scaling between
 space and time. 

 From rate equations such as (\ref{law}), one can simply work out the --- $d$-independent --- values of the critical
 exponents at mean-field level 
  (for instance for DP $\beta =1$, $\nu_\perp= 1/2$ and $z=2$  \cite{hinrichsen00}).

\subsection*{Field theory}

 The mean-field picture holds as long as the density remains homogeneous --- short-range correlated --- enough in the system so that
 the role of space and time fluctuations is indeed negligible, which is generally justified in high  dimensions.
 However, the reaction-diffusion processes under scrutiny are so-called `diffusion-limited', 
  which means that the diffusion is not efficient 
  enough to compensate the effect of reactions that  locally alter the density distribution and hence
 invalidate the
 mean-field approximation. A finer analysis therefore requires a systematic incorporation 
 of spacio-temporal  fluctuations, which can be achieved 
 through the construction of a field theory.

 A field theory to describe reaction-diffusion processes can be derived following a well-known formalism 
 due to Doi and Peliti \cite{doi76b}. The starting point is the master equation,
  which describes the time evolution of the probability distribution
  $P(\{n_i\},t)$ of the occupation numbers $n_i$ of the sites $i$ of a lattice 
  --- assuming no occupation restriction.
   The idea is to write the change in the occupation numbers on each site by means of
  creation and annihilation operators. Upon introducing   Fock states $|\{n_i\}\rangle$ 
 to represent the configurations of the lattice,  and 
 a formal state vector $|\Phi(t)\rangle =  \sum P(\{n_i\},t) |\{n_i\}\rangle$,
  the master equation can be recast into a Schr\"odinger-like equation for the state vector 
 $\partial_t |\Phi(t)\rangle=  - H |\Phi(t)\rangle$. Then, 
 resorting to  coherent-state path integrals  as in quantum mechanics, 
 one can construct a functional integral ${\cal Z}[\phi,\hphi] = \int {\cal D}\phi {\cal D}i \hphi \exp(-{\cal S}[\phi,\hphi])$, which captures exactly (up to the continuum space limit)
 the microscopic stochastic fluctuations \cite{doi76b,lee94}. 
 The time-dependent statistical averages of observables --- 
 which are necessarily mere functions of the occupation numbers --- 
  can then be computed from ${\cal Z}$.
 For the general processes  $A\xrightarrow{\sigma_m} (m+1)A$ and  $kA\xrightarrow{\lambda_k} \varnothing$, this procedure yields \cite{cardy96}:
\begin{multline}
 {\cal S}[\phi,\hphi]  = {\displaystyle \int}  d^d x\, d t\, 
 \Big\{ \hphi(x,t)\,\big(\pdt  - D\, \nabla^2\big)\,\phi(x,t) \\
- \lambda_k\, \big(1-\hphi(x,t)^k\big)\,\phi(x,t)^k
 +\sigma_m\,\big(1-\hphi(x,t)^m\big)\,\hphi(x,t)\,\phi(x,t)\, \Big\}. \label{actioncont}
\end{multline}
The diffusion is encoded in  the kinetic part and stands as  the Gaussian (quadratic) theory
 which corresponds to the Brownian motion. All the reactions give rise to potential interaction terms.
 The action (\ref{actioncont}) can then root field theoretical treatments,
   and in particular NPRG methods. This action indeed underlies in the following the investigation
  of various reaction-diffusion processes, 
 namely DP and BARW models. Before presenting these analyses (in sections \ref{secIV} and \ref{secV}), 
  we first give a brief overview of the NPRG techniques. 
  
\section{Non-perturbative renormalisation group out of equilibrium}
\label{secIII}

We do not intend to detail here the implementation of the NPRG, but rather set out its principle (and refer to \cite{berges02,bagnuls01} for reviews).
This formalism relies on Wilson's RG idea \cite{wilson74}, which consists in building  a sequence of scale-dependent effective models,
that interpolate smoothly between the short-distance physics at the
(microscopic) scale $k=\Lambda$ and the long-distance physics at the scale $k=0$,
 through progressively averaging over fluctuations.
Rather than expressing --- as in the original Wilsonian formulation --- the flow of effective Hamiltonians for the slow modes,
one can work out the flow of effective `free energies' $\gk$ for the rapid ones, following \cite{tetradis94,berges02}. $\gk$ thus only includes fluctuation modes with momenta $|\bq| \ge k$. At the scale $k=\Lambda$, no  fluctuation is yet  taken into account
and $\Gamma_{\Lambda}$ coincides with the microscopic action ${\cal S}$, while at  $k=0$, all  fluctuations are integrated out and $\Gamma_{0}$ is the analogue of the Gibbs free energy $\Gamma$ at thermal equilibrium, in that it encompasses the long distance properties of the system.
To construct $\gk$, one needs to freeze the slow modes. This is achieved by adding a scale-dependent mass-like  term to the original action \cite{berges02,canet04a,canet04c}:
\begin{equation}
\Delta {\cal S}_k[\hphi,\phi] =\int_{\bx,t} \hphi(\bx,t)\,R_k(\nabla^2,\p_t)\,\phi(\bx,t),
\end{equation}
where the cutoff function $R_k$ behaves as $R_k(\bq^2,\omega)\sim k^2$ (in Fourier space) for small momenta $|\bq| \leq k$ --- so that the slow modes are decoupled --- and $R_k$ vanishes for large momenta $|\bq| \ge k$ --- so that the rapid ones remain unaltered.
The ``partition
functions'' ${\cal Z}_k[j,\tj]= \int {\cal D}\phi\, {\cal D}i \hphi\, \exp(- {\cal S}- \Delta {\cal S}_k +  \int j\,\phi +  \int \tj\,\hphi)$ become therefore
$k$-dependent.
$\gk$ is obtained through the Legendre transform of   
$\log {\cal Z}_k[j,\tj]$:
\begin{equation}
\Gamma_k[\psi,\hp] +\log {\cal Z}_k[j,\tj]= 
\int j \psi +\int \tj \hp -\Delta {\cal S}_k[\psi,\hp]
\label{gammak}
\end{equation}
and is a functional of the conjugate fields
$\psi=\delta \log {\cal Z}_k/\delta j$ and $\hp=\delta \log {\cal Z}_k/\delta
\tj$.
Note that the last term in Eq. (\ref{gammak})  ensures
that $\Gamma_k$ has the proper limit at $k=\Lambda$: $\Gamma_{k=\Lambda}\sim
{\cal S}$ \cite{berges02,canet04c}.  An exact functional differential equation governs
the RG flow of $\Gamma_k$ under an infinitesimal change of
the scale $s=\log(k/\Lambda)$
\cite{berges02,canet04a}:
\begin{equation}
\partial_s \Gamma_k = \frac{1}{2} {\rm Tr} \int_{q,\omega} \partial_s \hat{R}_k \left(\hat\Gamma_k^{(2)} + \hat{R}_k\right)^{-1},
\label{dkgam}
\end{equation}
where $\hat{R}_k$ is the symmetric, off-diagonal, $2\times 2$ matrix
of element $R_k$ (in Fourier space) and $\hat\Gamma_k^{(2)}[\psi,\hp]$ the $2\times
2$ matrix of second derivatives of $\Gamma_k$ with respect to  $\psi$ and
$\hp$. Obviously, Eq.~(\ref{dkgam}) cannot be solved exactly
and one usually devices an \anz for $\Gamma_k$.  However, as the approximations used do not rely on the smallness of a parameter, the approach remains \npt in essence. In particular,  
it is not confined to weak-coupling regimes or to the vicinity of critical dimensions and  is therefore suitable to overcome the usual perturbative RG schemes.\\

Since the critical physics corresponds to the long distance ($\bq\to 0$) and long time ($\omega \to 0$) limit,
a sensible truncation  consists  in
expanding  $\Gamma_k$ in powers of  gradients \cite{tetradis94} and time derivatives.
Retaining only  the leading order in derivatives,  the most general \anz for $\Gamma_k$ related to the field theory (\ref{actioncont}) 
 writes \cite{canet04a}:
\begin{equation}
\gk(\psi,\hp)=\int d^d\bx \, d t\, \Big\{\hp(\bx,t)\, \big[D_k[\psi,\hp] \p_t - Z_k[\psi,\hp] \nabla^2\big]\psi + U_k[\psi,\hp].
\label{gamcomkpz}
\end{equation}
The effective potential $U_k$ encompasses all possible reactions. The kinetic renormalisation
 functions $D_k$ and $Z_k$ of the diffusive part
 account for the  anomalous scalings of the fields and of time.
 Indeed, the anomalous dimension $\eta$ of the fields and the dynamic exponent $z$
 are defined such that at criticality,
 $\psi\hp \sim k^{d+\eta}$ and $\omega\sim k^z$ respectively \cite{wijland98}. 
 It follows that $\eta_k=-\p_s \ln D_k$ and  $z_k= 2 + \eta_k + \p_s \ln Z_k$, the critical exponents
 $\eta$ and $z$ corresponding to the  ($k$-independent) fixed point values of $\eta_k$ and $z_k$.

 The flow equations for the potential $U_k$ and the renormalisation
 functions $Z_k$ and $D_k$ have been established for generic  reaction-diffusion processes in \cite{canet04a}.  
 Different levels of  
 approximation can be implemented.
  The simplest one, the so-called local potential approximation (LPA), consists in neglecting
  the kinetic renormalisations, {\it i.e.} in setting $Z_k=D_k=1$, upon which $z=2$ and $\eta = 0$.
 The LPA generally enables one to get a reliable qualitative picture 
 of the physics as well as a fairly accurate
 determination of the static exponent $\nu$.
  This approximation can be refined by including  (field-independent) kinetic renormalisation
  coefficients $D_k$ and $Z_k$, which allows for non-trivial
 estimates of $\eta$ and $z$. This approximation is  referred to 
 as leading order (LO). Finally, to get more accurate values of the exponents requires to incorporate 
 the field dependence of the kinetic functions $Z_k[\psi,\hp]$ and $D_k[\psi,\hp]$ 
 --- referred to as next to leading order (NLO) ---
  which becomes much more tedious numerically.

In order to study a specific model, the generic flow equations $\p_s U_k$, $\p_s Z_k$ and $\p_s D_k$ must be solved 
 for the functionals $\gk$ respecting the symmetries of the problem,
  starting at scale $\Lambda$ from the corresponding
  microscopic action (\ref{actioncont}).
 We  focus in the next section on the DP universality class.

\section{Directed Percolation}
\label{secIV}

The action corresponding to the DP processes (\ref{DPrules}) can be 
 deduced from (\ref{actioncont}) and writes:
\begin{equation}
S[\phi,\hphi]= \int_{x,t}\! \hphi (\partial_t\!  - D\!
\nabla^2)\phi - \lambda (1\!-\!\hphi^2)\phi^2 + \sigma
(1\!-\!\hphi)\phi \hphi -\mu\!(1\!-\!\hphi)\phi.
\end{equation}
 It  can be conveniently rewritten  expanding the response field around 
 its stationary solution  $\hphi(x,t) = 1+\tphi(x,t)$ and rescaling the fields according to  
 $\tphi\to\sqrt{2\lambda /{\sigma}}\,\,\tphi$ and $\phi\to\sqrt{{\sigma}/{2\lambda}}\,\,\phi$, which leads to:
\begin{equation}
 {\cal S}_{DP}[\phi,\tphi]  = {\displaystyle \int}  d^d x\, d t\, 
 \Big\{ \tphi\,\big[\pdt  - D\, \nabla^2 - (\sigma-\mu)\big]\,\phi
    + \sqrt{2\,\sigma\,\lambda}\,\big[\tphi\,\phi^2 -
\tphi^2\,\phi\big] + \lambda\,(\phi\tphi)^2 \Big\}. \label{actionDP}
\end{equation}
 This action turns out to be equivalent to the so-called Reggeon field theory \cite{cardy80}, which has been studied in 
 particle physics since the early seventies \cite{moshe78}.
 The critical exponents have been computed in that context
  to two-loop order.
 However, the upper critical dimension of this field theory is $d_c=4$\footnote{Note that in this paper, $d$ denotes the dimension of {\it space} --- {\it i.e.} does not include the time direction.}, which is distant from the physical 
 dimensions  $d=1$ and $d=2$. Moreover, in spite of its simplicity, the DP model has no exact
 solution in $d=1$ --- at variance with the Ising model. Thus, the best estimates 
 of the critical exponents of DP rely on numerical calculations and are given in table \ref{tab}.

  Providing  analytical estimates of the DP exponents thus requires a method that is not confined 
 to the vicinity of a critical dimension. This motivates the use of a \npt approach 
 --- the NPRG --- to fulfil this task,
  since the latter allows to span  arbitrary dimensions.
  The generic flow equations introduced in 
 section \ref{secIII} can be used to compute the critical exponents of the DP universality class,
 provided  the symmetries of the DP model are specified. The bare 
 action (\ref{actionDP}) turns out to be invariant
 under the change:
\begin{equation}
\left\{
\begin{array}{r l l}
\phi(x,t)  & \to & -\tphi(x,-t) \\
\tphi(x,t) & \to & -\phi(x,-t),
\end{array}
\right. \label{defsymrap}
\end{equation}
which is called the `rapidity' symmetry.
The effective potential $U_k(\psi,\tp)$ and kinetic renormalisation functions
   $D_k(\psi,\tp)$  and  $Z_k(\psi,\tp)$ --- denoted $X_k$ in the following ---
 must be invariant under the rapidity transformation (\ref{defsymrap}). 
 This in turn  imposes that the generic term of the Taylor expansion of the $X_k$'s  be of the form
 $a_{\alpha\beta}(\psi^\alpha\tp^\beta + (-1)^{\alpha+\beta}\tp^\alpha\psi^\beta)$,
 which only involves the two invariant combinations $x = \psi\tp$ and $y =\psi-\tp$.
 Thus,  parameterising the $X_k$'s in terms of the invariants $x$ and $y$ 
 ensures that they satisfy the rapidity symmetry constraints. 

 The flow equations for $U_k$, $D_k$ and $Z_k$ then have to be solved numerically.  
 One can integrate the flows with the scale $s$ starting from some initial bare rates \cite{canet04a}.
 For a fine tuned initial mass $\Delta$,
   the (dimensionless) effective potential 
  flows to a fixed function, which corresponds to criticality.
 The exponents are calculated in the vicinity of this fixed solution. 
 This procedure has been performed at different levels of approximation, namely the LPA, the LO 
 and the NLO \cite{canet04a,canet04c}.  
  The results are displayed in table \ref{tab} and show that the estimates converge fairly rapidly
   as the approximation is enriched to  values in good agreement with the best numerical estimates.
\Table{\label{tab} 
 Critical exponents of the DP universality class, from NPRG calculations within different
 levels of approximation --- LPA, LO in $d=1,2,3$  \cite{canet04a} and NLO in $d=2,3$ \cite{canet04c}. 
  The last column gathers  the best numerical estimates, ensuing from Monte Carlo (MC) simulations \cite{jensen92,voigt97} and series expansions \cite{jensen99}.\\} 
\begin{tabular}{ c c c c c c }
\br
\ \ \  d \ \ \  &\quad \quad &\ \ LPA \cite{canet04a} \ \  &  LO\cite{canet04a}  &  \ \ NLO \cite{canet04c}  \ \  &\ \  MC+series \cite{jensen92,voigt97,jensen99}  \ \ \\ \br
    &${ \nu}$    & 0.584 & 0.548 &{ \it 0.59}  &0.581(5)\\
3   &${ \beta}$  & 0.872 & 0.782 &{ \it 0.83}  &0.81(1)\\
    &${ z}$      & 2     & 1.909 &{ \it 1.90}  &1.90(1)\\ \mr
    &${ \nu}$    & 0.730 & 0.623 &{ \it 0.73}  &0.734(4)\\
2   &${ \beta}$  & 0.730 & 0.597 &{ \it 0.59}   &0.584(4)\\
    &${ z}$      & 2     & 1.884 &{ \it 1.70}  &1.76(3)\\ \mr
    &${ \nu}$    & 1.056 & 0.888 &    &1.096854(4)\\
1   &${ \beta}$  & 0.528 & 0.505 &    &0.276486(8)\\
    &${ z}$      & 2     & 1.899 &    &1.580745(10)\\ \br
\end{tabular}
\endTable

\section{Branching and annihilating random walks}
\label{secV}

Initially introduced by Bramson and Gray \cite{bramson85}, models of BARW can be seen as 
 reaction-diffusion processes endowed with the generic  rules:
\begin{equation}
 A \xrightarrow{\sigma_m} (m+1) A, \hspace{1cm} k A \xrightarrow{\lambda_k} \varnothing,  \label{barw}
\end{equation}
 from which is excluded the  spontaneous decay $A\to\varnothing$  ({\it i.e.} $k>1$).
 The particles can therefore only
 disappear through a $k$-body annihilation. This restriction has drastic consequences
  since at mean-field level, it forces
  the system to reach  an active state with mean density $n_s=[(m\, \sigma_m)/(k\, \lambda_k)]^{1/(k-1)}$
   for any non-zero branching rate $\sigma_m$. A trivial transition occurs at $\sigma_m=0$, where the model coincides 
 with the pure annihilation model. The latter is well-controlled theoretically \cite{lee94}
  and it predicts an algebraic
 decay of the density, which follows $n(t) \sim t^{-1/(k-1)}$ 
 above the upper critical dimension $d_c(k) = 2/(k-1)$ and  is slowed down by fluctuations 
  below $d_c(k)$ \cite{lee94}. However, early simulations \cite{grassberger84,takayasu92} have evidenced  in low dimensions
 the existence of  absorbing phase transitions at non-vanishing branching rates
 in these systems. These transitions
 have been found to  belong to two different universality 
 classes, depending on the parity of $m$ and $k$ \footnote{Note that it has been argued in \cite{kockelkoren03}
 that the parity conservation had no influence on similar reaction-diffusion systems and that
 the parity-wise distinction between odd and even BARW processes could be somewhat incidental, see also \cite{dornic01}.}.
  The corresponding models 
 are  therefore called `odd' and `even' BARW.
 Obviously, the effect of fluctuations cannot be neglected in low dimensions and need to be incorporated
 in a systematic way. In a seminal paper \cite{cardy96}, Cardy and T\"auber have derived the complete field theory 
 for BARW and analysed both `odd' and `even' categories  through perturbative RG. Their main results are
  summarised in the following. 
 We mention here a straightforward outcome of their analysis, which is that for given $k$ and $m$, 
 all the reactions $k-2,k-4\dots$ and $m-2,m-4\dots$ are generated 
 under renormalisation,  the lowest $m$ and $k$ processes standing as the most relevant ones. Thus,  investigating
  the generic processes ($A\xrightarrow{\sigma}2A,2A\xrightarrow{\lambda}\varnothing$)
 and ($A\xrightarrow{\sigma}3A,2A\xrightarrow{\lambda}\varnothing$) suffices 
 to describe the  critical behaviour of the odd and even BARW respectively. In both cases,
 the general idea advocated in \cite{cardy96} is to start from the pure pair
  annihilation process in the vicinity of its upper
 critical dimension $d_c=2$ and  to treat perturbatively a small branching rate $\sigma\sim \epsilon$.
 The problem  is then to determine if fluctuations irremediably destabilise
  the absorbing state --- as suggested by mean-field ---
  or if they allow for a non-trivial transition to take place.
  We first focus on the case of odd-BARW.

\subsection{odd-BARW: phase diagram}
\label{subsecVI}

\begin{figure}[ht]
\caption{Sketch of the transition lines of the BARW 
 $A \xrightarrow{\sigma} 2A$, $2A\xrightarrow{\lambda}\varnothing$ computed in \cite{cardy96}. 
 For each dimension, the active phase lies above the transition 
line, the absorbing phase below.\\}
\begin{indented}
\item[]\includegraphics[width=55mm,height=46mm]{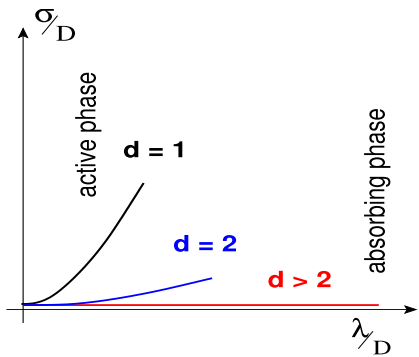}
\end{indented}
\label{schem}
\end{figure}

For odd-BARW, the spontaneous decay $m = -1$ turns out to be generated under renormalisation
 through the combination $A\to2A\to\varnothing$ at a renormalised rate $\mu_R$ depending
 on the bare rates $\lambda$ and $\sigma$. The mass $\Delta_R = \sigma_R -\mu_R$
  hence acquires corrections  and  
  the aim is to determine whether it can become negative for a non-zero bare $\sigma$ and drive
 a transition to an absorbing state.
 If so,
  the transition will naturally belong to the DP class since then the renormalised action
 is identical to (\ref{actionDP}).
 Cardy and T\"auber computed $\Delta_R$ following two different procedures 
 which yielded consistent results  \cite{cardy96} --- that are sketched 
 on  figure \ref{schem}. 
  The conclusion is the following: fluctuations 
 are strong enough to induce a non-trivial phase transition only in $d \leq 2$.
 The transition line is given by $\sigma_c = D [\lambda/(2 D \pi \epsilon)]^{2/\epsilon}$ 
 for $d<2$ and is exponentially suppressed following 
  $\sigma_c = D \exp(-4\pi D/ \lambda)$ in $d=2$ \cite{cardy96}.
 The perturbation theory breaks down above two dimensions. However, since
  the annihilation rate becomes irrelevant in $d>2$ and since moreover directed  random walks are known to
 be reentrant only for transverse dimensions $d<2$ 
 (that is the probability of intersection of two random paths vanishes above
 two transverse dimensions),
  Cardy and T\"auber inferred that an absorbing state could no longer exist above two 
 spatial dimensions \cite{cardy96}. In other words, their analysis  suggests that the mean-field results
  are recovered for $d>2$, that is the system 
 is always in an active state above two dimensions. \\

 We re-examined odd-BARW using NPRG. It should be stressed that critical rates, like critical
 temperatures, are non-universal quantities. Since the NPRG formalism allows to keep track, through 
 the scale integration of
 the flow, of the initial microscopic (bare) action, it enables one to determine the critical
 rates for which the flow leads to a scale-invariant effective potential. We thus integrated the flow 
 equations $\p_s U_k$,  $\p_s Z_k$ and  $\p_s D_k$ at LO for different initial 
 bare rates in dimensions 1 to 6 \cite{canet04b}.
\begin{figure}[ht]
\caption{Phase diagrams of the BARW 
$A \xrightarrow{\sigma} 2A$, $2A\xrightarrow{\lambda}\varnothing$
in dimensions 1 to 6, from \cite{canet04b}. Lines present NPRG results, rescaled as explained 
in the text. Symbols follow from numerical simulations. 
For each dimension, the active phase lies on the left of the transition 
line, the absorbing phase on the right.\\}
\begin{indented}
\item[]\includegraphics[width=60mm,height=90mm,angle=-90]{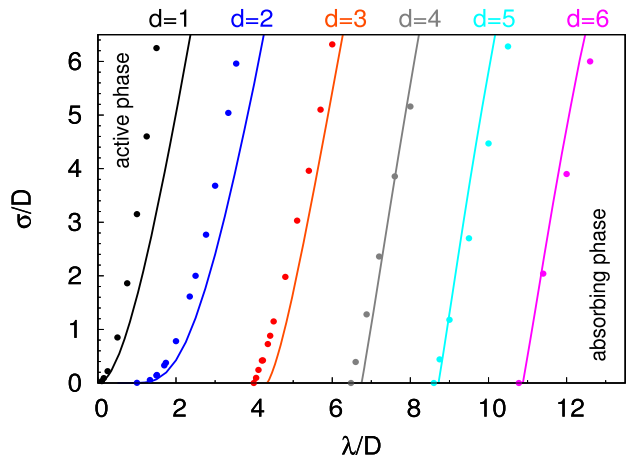}
\end{indented}
\label{diag}
\end{figure}

  We also performed extensive numerical simulations, using an efficient algorithm 
  introduced in \cite{kockelkoren03} to back up our findings.
  For given values of $\lambda/D$, we generated  on a lattice 
 the stochastic time evolution of particles subject to  the dynamics
   $A \xrightarrow{\sigma} 2A$, $2A\xrightarrow{\lambda}\varnothing$
  and we searched for the critical rates  $\sigma/D$ 
  which yield an algebraic decay of the density --- in dimensions 1 to 6  \cite{canet04b}. 

  Prior to  comparing  the obtained `discrete' critical points ($\lambda/D,\sigma/D$) with the `analytical' ones ensuing from NPRG,
  the latters need 
   to be rescaled  by dimensional factors ($C^{2-d},C^2$) 
  to account for the continuous space limit underlying the field theory \cite{canet04b}.
  We emphasise that, however,
  matching a single point 
  suffices to uniquely fix the value of the $C$ parameter for all dimensions.
 Both numerical and (rescaled) analytical  
 phase diagrams are displayed on figure \ref{diag} and show a remarkable agreement.
 Indeed, let us recall that,
  at variance with universal quantities such as critical exponents, non-universal ones 
 depend on all irrelevant operators (and microscopic details) 
 and the accuracy of NPRG results was therefore unexpected.

\begin{figure}[ht]
\caption{Evolution of the thresholds $(\lambda/D)_{\hbox{\footnotesize th}}$ with the dimension,from
  \cite{canet04b}.\\}
\begin{indented}
\item[]\includegraphics[width=40mm,height=90mm,angle=-90]{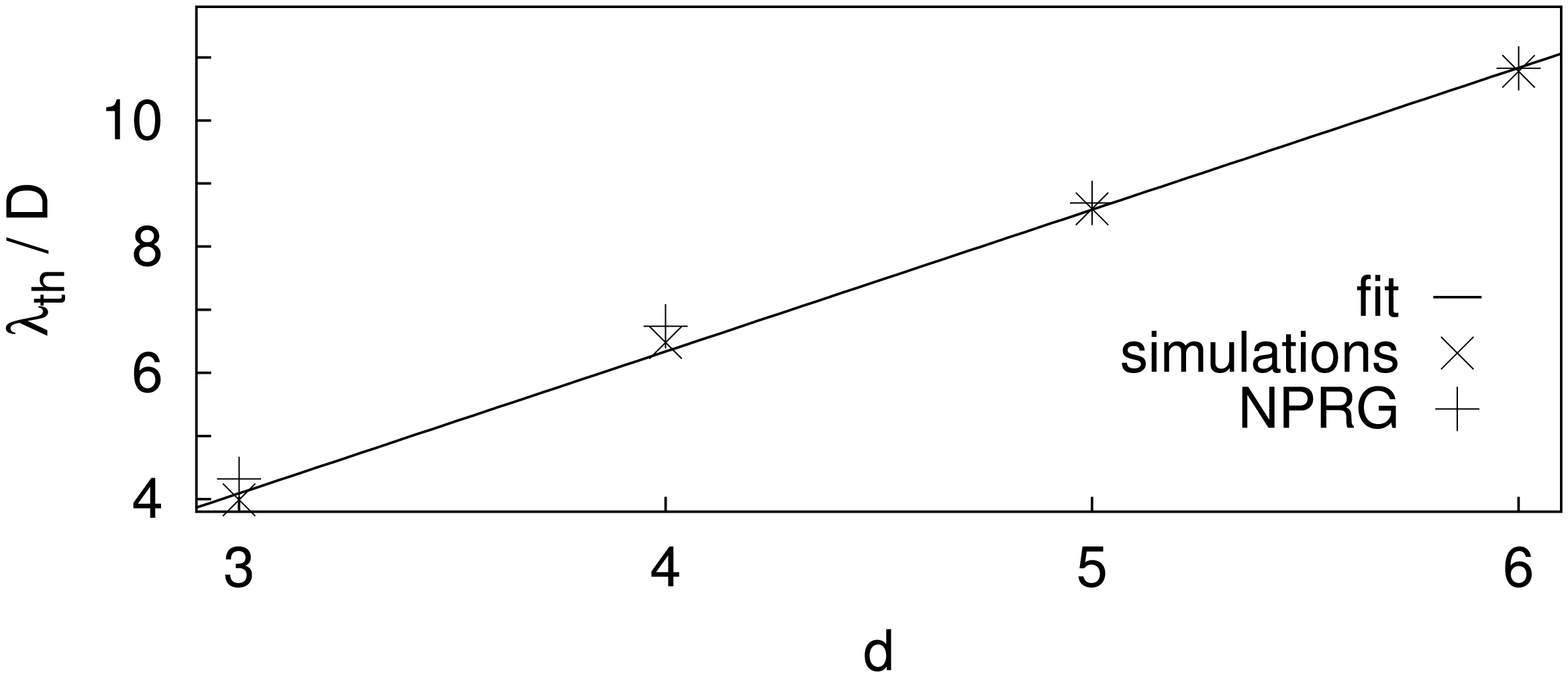}
\end{indented}
\label{seuil}
\end{figure}

 Let us now analyse these phase diagrams. First,
 the transition lines  of figure \ref{diag} are in accordance with the perturbative results in their region
 of validity, that is in the vicinity of the origin. Both numerical and analytical transition lines
  are indeed quadratic in $d=1$
 and exponential in $d=2$ \cite{canet04b}. But away from the origin,
  the phase diagrams we obtained drastically differ from the perturbative results  \cite{cardy96}
 sketched on figure \ref{schem}.
  Indeed, we found a transition  in $d=3$ and 
 in fact in all probed dimensions up to $d=6$. The inactive phase appears to emerge only after a finite
 threshold $(\lambda/D)_{\hbox{\footnotesize th}}$ which is  in essence a \npt feature.
  It is indeed out of reach of any
 perturbative expansion around the origin. Furthermore, the transition lines seem to undergo a simple
 drift as the dimension grows. A closer investigation of the variation of the thresholds
 with the dimension reveals  that they grow linearly with $d$ (which has 
 been checked up to $d=10$) as shown on figure \ref{seuil}. 
 This strongly suggests that an absorbing DP transition occurs in all finite dimensions, or in other words
 that the mean-field result is recovered only in the $d\to\infty$ limit \cite{canet04b}.\\
 
  We emphasise that this study unveils  a remarkable instance where fluctuations
  turn out to {\it qualitatively} invalidate 
  the mean-field and even one-loop phase diagrams. Indeed, fluctuations
  do not only bring quantitative corrections 
 to the critical rates but build up genuinely \npt effects that affect 
 the very existence of the transition.
We now come to the case of even-BARW and  show that \npt features are even more crucial there since
 they entirely entail the physics of the model.

\subsection{even-BARW: universality class}
\label{subsecVII}

The models of even-BARW have concentrated much attention \cite{grassberger84,cardy96,dornic01}
  as they embody the first example of an absorbing
 phase transition {\it not} belonging to the DP class, but to a new one --- improperly, see $^\ddagger$ ---
  called `PC' for parity conserving.
  For these models, a phase transition has been observed numerically in one dimension and characterised by 
 non-DP exponents. To provide an  analytical support to these findings,
 Cardy and T\"auber have attempted a  RG treatment of the field theory of even-BARW,
  which writes \cite{cardy96}:
\begin{equation}
S[\phi,\hphi]= \int_{x,t}\! \hphi (\partial_t\! -\! D
\nabla^2)\phi - \lambda (1\!-\!\hphi^2)\phi^2 + \sigma
(1\!-\!\hphi^2)\phi \hphi.
\label{bareaction}
\end{equation}
 
 Their analysis showed the appearance of a second upper critical dimension $d_c\simeq 4/3$ below
 which the branching rate $\sigma$
  becomes irrelevant, allowing for the annihilation fixed point $F_A$ to 
  become
 stable and possibly root an absorbing state in $d=1$. 
 Furthermore,  in a direct calculation in $d=1$, Cardy and T\"auber \cite{cardy96} managed to identify a combination of $\sigma$ and $\lambda$ that admits a fixed point at one-loop
 order. However, it yielded 
  poorly determined exponents and the extension of this calculation to higher orders appeared
 problematic \cite{cardy96}. \\

 We therefore re-analysed the even-BARW model by means of NPRG methods \cite{canet05}. 
 We once again exploited the generic flow equations derived in section \ref{secIII}, 
 specified for even-BARW theory.
 The action (\ref{bareaction}) is $Z_2$ symmetric. The effective potential $U_k$
 should hence only depend on the quadratic invariants
 $\psi^2$, $\hp^2$ and $\hp\psi$,  structure which appears to be preserved by the flow equations
 $\p_s X_k$'s. We here consider 
  the LPA, which suffices to obtain a non-trivial picture of the physics of the model. 
 The effective potential is Taylor expanded around the stationary solution $(\hp=1, \psi=0)$ and truncated
 at a given power $n$ of the fields.
\begin{figure}[ht]
\caption{Variation with $d$ of the eigenvalues 
of the fixed points in the lowest-order LPA, from \cite{canet05}. Blue lines: pure annihilation
fixed point $F_{A}$. Red lines: non-perturbative fixed point $F_{PC}$  
whose eigenvalues are both negative for $\frac{4}{3}<d<1.3784\ldots$ 
and complex-conjugated at larger $d$ (only the real part is plotted).\\}
\begin{indented}
\item[]\includegraphics[width=60mm,height=90mm,angle=-90]{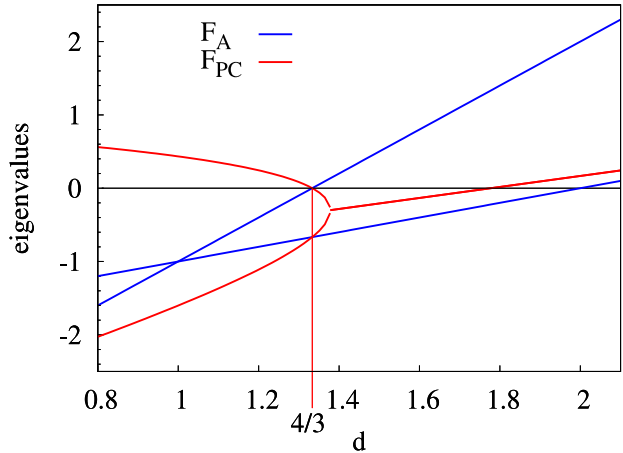}
\end{indented}
\label{vp}
\end{figure}

\begin{figure}[ht]
\caption{Flow diagram of the lowest-order LPA in $d=1$ (arrows
represent the RG trajectories as $s$ is decreased towards the ``infra-red'',
macroscopic
 limit  $s \to -\infty$), from \cite{canet05}.\\}
\begin{indented}
\item[]\includegraphics[width=50mm,height=90mm,angle=-90]{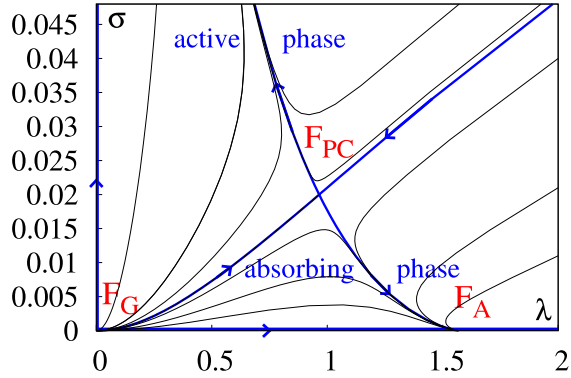}
\end{indented}
\label{flot}
\end{figure}

  Even from the lowest order $n=2$ (corresponding to $U_k= - \lambda_k (1-\hp^2)\psi^2 + \sigma_k
(1-\hp^2)\psi \hp$), the flow equations for the coupling constants $\lambda_k$ and $\sigma_k$
 exhibit, in addition to the Gaussian fixed point $F_G$ = \{$\sigma=0,\lambda=0$\} and to the pure annihilation
 fixed point $F_A$ = \{$\sigma=0,\lambda_A \neq 0$\}, a \npt solution 
 $F_{PC}$ = \{$\sigma_{PC} \neq 0,\lambda_{PC} \neq 0$\} \cite{canet05}.
  The latter governs an absorbing
 transition in dimension $d<4/3$. It becomes unphysical 
(with a negative $\sigma_{PC}$) and thus plays 
 no role for $d>4/3$. Indeed, the stability of  $F_{PC}$ and $F_A$ can be read
 off from  figure \ref{vp} which displays their eigenvalues. 
For $d>4/3$, the annihilation fixed point is the only physical one and it is once unstable (in the $\sigma$ 
 direction), implying that the system is always active. When $d$ decreases, $\sigma_{PC}$ approaches zero
 until   $F_{PC}$ crosses $F_A$ in $d=4/3$, where
 they exchange stability.  Below $d=4/3$, $F_A$ is hence stable,
  describing the absorbing phase and  $F_{PC}$ acquires
 an unstable direction, thus driving  a phase transition \cite{canet05}. 
 The corresponding flow diagram  is depicted for $d=1$ in figure
 \ref{flot}. 

The critical exponent $\nu$ can be determined in the vicinity of $F_{PC}$ and it converges
 as the order $n$ of the field truncation is increased to $\nu = 2.0\pm 0.1$, which is already in fair
 agreement with Monte Carlo simulations yielding $\nu=1.85\pm 0.1$.
 One would then need to go to the next order
 in the derivative expansion and include non-trivial kinetic renormalisations analogous to $Z_k$ and $D_k$
 to get a determination of the other exponents. 
We stress that $F_{PC}$ does not 
 become Gaussian in any dimension and is thus genuinely \npt. This explains why it cannot be reached
 {\it via} any perturbative expansion.

\section{Conclusion and prospects}
Through this review we have advocated the use of a \npt field-theoretic method, the NPRG, to investigate 
 \nequ systems. We have highlighted the valuable results that this method has allowed to provide
 for various reaction-diffusion processes. Most importantly, we have unveiled that \npt effects turn out
 to play a crucial role in the physics of these models. Indeed,  we have
 first computed the phase diagram of odd-BARW
 and evidenced the existence of a \npt threshold for the absorbing phase to emerge above two dimensions.
 This very threshold explains the failure of perturbative treatments which incorrectly lead to the conclusion
 that the transition disappears for $d>2$. Furthermore, we have studied the NPRG flow equations of the even-BARW
 model and found a genuine \npt fixed point --- {\it i.e} non-Gaussian in any dimension --- which
 is responsible for the PC absorbing phase transition in low dimensions  and becomes unphysical
 above $d=4/3$. It hence provides 
 a theoretical back-up for the transition observed in numerical simulations, 
   reproducing in particular the numerical results for the $\nu$ exponent in $d=1$.\\

The NPRG appears as an efficient tool to tackle \nequ systems, which opens many prospects.
 First, it would allow to investigate critical properties of other reaction-diffusion processes.
  For instance, the universality class of the so-called pair contact process with diffusion model has been the
 subject of a long-lasting debate \cite{henkel04}.
  The influence of quenched disorder in DP models is also of great 
 interest since the latter is suspected to hinder experimental realisations
  of this universality class \cite{hinrichsen00}.
  Beyond the scope of reaction-diffusion processes, this technique 
 has been applied to the study of the   roughening transition in surface growth,
   generically modelled by the notorious Kardar-Parisi-Zhang
 equation \cite{kardar86}. It has allowed to obtain non-trivial results \cite{canet05b} and it would be of utmost interest 
  to push further the investigation of growth phenomena.
  The NPRG methods could finally allow to give some insights into glassy dynamics.

\ack

The author is deeply indebted to H. Chat\'e, B. Delamotte, O. Deloubri\`ere, I. Dornic, M. Mu\~noz, and N. Wschebor who collaborated 
 on the different works reviewed here. The author also wishes to thank the organisers of the conference RG2005 where this talk was presented.
 This work benefits from the financial support by the European Community's Human Potential Programme under contract HPRN-CT-2002-00307, DYGLAGEMEM.


\end{document}